\newlength\smallfigwidth
\newlength\tinyfigwidth
\documentclass[aps,prb,twocolumn,floatfix,amsmath,amssymb,showpacs,showkeys]{revtex4}

\newcommand{\be}{\begin{equation}}
\newcommand{\ee}{\end{equation}}
\newcommand{\ba}{\begin{align}}
\newcommand{\ea}{\end{align}}
\newcommand{\bn}{\begin{eqnarray}}
\newcommand{\en}{\end{eqnarray}}

\usepackage{graphicx}
\begin{document}

%% \begin{center}
%% {\Large \bf Magnetic anisotropy of elongated thin nano-islands}
%% 
%% \vskip 0.2cm
%% G.\ M.\  Wysin$^{*}$, W.\  A.\ Moura-Melo, L.A.S. M\'ol and A.\ R.\ Pereira  \\
%% \vskip 0.2cm
%% Departamento de F\'isica \\
%% Universidade Federal de Vi\c cosa, Vi\c cosa 36570-000, Minas Gerais, Brazil \\
%% December, 2011
%% \vskip 0.1cm
%% $^{*}$wysin@phys.ksu.edu,
%% http://www.phys.ksu.edu/personal/wysin \\
%% Department of Physics, Kansas State University,
                %% Manhattan, KS 66506-2601 \\
%% \vskip 0.2in
%% \end{center}

\preprint{KSU-UFV-Wysin et al.}

\title{Magnetic anisotropy of elongated thin ferromagnetic nano-islands for
    artificial spin ice arrays}
\author{G.\ M.\  Wysin}
\email{wysin@phys.ksu.edu}
\homepage{http://www.phys.ksu.edu/personal/wysin}
\affiliation{Department of Physics, Kansas State University, Manhattan, KS 66506-2601 }
\author{W.\  A.\ Moura-Melo}
\email{winder@ufv.br}
\author{L.A.S. M\'ol}
\email{lucasmol@ufv.br}
\author{A.\ R.\ Pereira}
\email{apereira@ufv.br}
\affiliation{Departamento de F\'isica,
Universidade Federal de Vi\c cosa, Vi\c cosa 36570-000, Minas Gerais, Brazil}
\date{December 7, 2011}
\begin{abstract}
{ The energetics of thin elongated ferromagnetic nano-islands is considered for
some different shapes, aspect ratios, and applied magnetic field directions.  These
nano-island particles are important for artificial spin-ice materials.   For low 
temperature, the magnetic internal energy of an individual particle is evaluated numerically 
as a function of the direction of a particle's net magnetization.  This leads to  
estimations of effective anisotropy constants for (1) the easy axis along the particle's long
direction, and (2) the hard axis along the particle's thin direction.  A spin relaxation algorithm 
together with fast Fourier transform for the demagnetization field is used to solve the
micromagnetics problem for a thin system.   The magnetic hysteresis is also found.  The 
results indicate some possibilities for controlling the equilibrium and dynamics in spin-ice 
materials by using different island geometries.}
\end{abstract}

\maketitle

%-------------------------------------------------------------------
\section{Introduction: Elongated thin ferromagnetic nano-islands}
%-------------------------------------------------------------------
Disordered and frustrated magnetic states such as those present in artificial spin 
ices \cite{Wang06,Remhof08} continue to draw interest, due to their competing ground states,
magnetic monopole excitations \cite{Castel08}, string excitations \cite{Mol09,Mol10,Morgan11,Moller09} 
and the difficulty to achieve thermal equilibrium.    Those systems are composed from elongated 
magnetic islands or particles of some length $L_x$ (several hundred nanometers) and width $L_y$ 
grown or etched lithographically to a small height $L_z$ on a substrate, whose geometric demagnetization 
effects (effectively, internal dipolar interactions) lead to a strong magnetic anisotropy.   
The typical islands have $L_z$ much less than $L_x$ or $L_y$.   Obviously any very thin magnet 
acquires an effective easy-plane anisotropy \cite{Gioia97}, and if the particle is narrow as well, 
the long direction becomes an easy axis.  The demagnetization field within an individual particle 
is responsible for this, making the plane of the island ($xy$-plane) an easy plane, and the 
$x$-axis an easy axis.  Then net magnetic moment $\vec\mu$ acts somewhat 
like an Ising variable with a defined easy axis $\hat{x}$. These islands are arranged 
into ordered arrays to produce, for example, square lattice or kagome lattice artificial spin-ices.
The analysis of spin ice models assumes that such particles have only the two states with  $\vec\mu$ 
either aligned or anti-aligned to the particle's easy axis.  The dipolar interaction between different
particles on one of the spin-ice lattices leads to the ice-rules, such as the ``two in / two out'' rule for
the square lattice and pyrochlore spin ices \cite{Castel08}.  Such ice rules are only energetic {\em preferences}, 
however, and only indicate the preferred states of the magnetic moments.  They are not absolute rigid 
statements about the allowed states.    Thus, the intention here is to investigate the energetics of 
the  fluctuations away from this Ising aligned state, in the individual elliptical islands that are
used to compose a spin-ice system.

At some level, there must be transitions between these Ising-like states.   
An individual particle may contain thousands of atomic spins, leading 
to a substantial energy barrier that must be surpassed to flip the Ising state of a particle.  
Hence, the dynamics is greatly constrained by such energy barriers.  It is our interest here 
to discuss how this barrier depends on the particular geometry of the islands, and make some 
evaluations of the dependence of the effective potential on the island shape and height.  
The types of shapes we consider are ellipses.  Thin single domain ellipses
were studied by Wei {\it et al.} \cite{Wei03}, who found that the reversal process involves
close to a uniform Stoner-Wohlfarth rotation, but with reduced energy barriers due to some
non-uniformity of the magnetization.  However, we find here that for high-aspect ratio ellipses,
this non-uniformity is minimal and a uniform rotation model could be very useful.

Although the theory for spin ice has been developed for Ising-like magnetic moments, their dynamics 
requires a different model.  In reality, the underlying magnetic moment must be evolving
from much more complex dynamics.  The reversal of an individual island, in the dipolar fields of its
surrounding islands, must be a complex process, and could involve the motion of domain walls and
vortices within the individual particles, or an impeded rotation of the local magnetization mostly in 
unison.  But in the assumption of strong ferromagnetic exchange inside a particular particle, and
a uniform externally applied field, one can investigate the reversal process using different approaches 
to the micromagnetics \cite{Suess06}, and see whether vortices or domain walls play any significant role. Especially, 
one can investigate whether there are intermediate metastable vortex or domain-wall states as steps of 
the reversal.   To a great extent for the thin elliptical particles considered here, the reversal proceeds mostly 
as a nearly uniform but impeded rotation of the magnetization of the particle \cite{Wei03},  although
the switching fields are reduced compared to a perfectly uniform rotation.  Hence, the idea 
of an Ising spin for a particle can be replaced by a three-dimensional magnetic moment $\vec\mu$, moving in 
some anisotropy potential, but free to point in any direction, if enough energy becomes available to it.

Obviously, by changing the aspect ratios $g_1\equiv L_x/L_y$ and $g_3\equiv L_x/L_z$ of the particle, its 
effective anisotropy changes.   The deviation of the ratio $L_x/L_y$  from $1$ determines the strength 
of an easy-axis anisotropy constant, call it $K_1$, for the net magnetic moment to rotate within the 
$xy$ plane.  The other aspect ratio of length to thickness, $L_x/L_z$, determines the difficulty for 
the magnetic moment $\vec\mu$ to tilt out of the $xy$ plane.  Thus it determines the strength of a hard
axis anisotropy constant, call it $K_3$.  The goal here is to make some accurate estimates for
these constants, and in the process, to justify a more generalized description of the magnetic
dynamics, not based on an Ising variable, but rather, on an effective three-dimensional magnetic
moment, that is allowed to make deviations from the Ising axis.  For a particle whose hard axis is
along $\hat{z}$ and easy axis is along $\hat{x}$, an effective potential that approximately 
represents their energies is shown to be
\be
\label{potential}
E = E_0+ K_1 \left[1-(\hat\mu \cdot \hat{x})^2\right] + K_3 (\hat\mu \cdot \hat{z})^2
\ee
where $\hat\mu$ is the unit vector pointing in the direction of the particle's net magnetic moment.
$E_0$ is the energy when the magnetic moment $\hat\mu$ is along the easy axis.
This type of potential is continuous, in contrast to the two-state Ising particle, having a well-defined
energy barrier, and having a more realistic dynamics.  Further, it will give the possibility of 
controlling the thermodynamics of spin-ices via changes or variations in the nano-island structure,
that can modify the energy barrier.

The calculational approach is a modification of usual micromagnetics \cite{Cervera99,Garcia03}, as follows.  
A particle is partitioned into cells of size $a\times a\times L_z$, under the assumption of the local 
magnetization $\vec{M}(\vec{r})$ being independent of the $z$-coordinate (along the thin dimension).   Thus, 
there is only a single layer of cells in the $xy$-plane, with the desired shape, say, an ellipse of major 
diameter $L_x$ and minor diameter $L_y<L_x$.   The saturated magnetization in each cell interacts
with the neighboring cells by ferromagnetic exchange, an externally applied magnetic field, and interacts 
with all cells via the demagnetization field.  The demagnetization field is calculated using an effective 
Green's function that applies for thin systems \cite{Huang03}, see below, with the calculation accelerated by
using a 2D fast Fourier transform (FFT).  To evolve towards the nearest (possibly meta-) stable 
magnetic state, we do not use integration of the Landau-Gilbert spin dynamics equations with
damping.  Instead, a faster procedure is to use a local spin-alignment algorithm, that involves
no damping parameter.   In one step of this algorithm, each cell's magnetic moment is pointed towards 
the local total magnetic field that is instantaneously producing a torque on that cell.  The same 
procedure is applied to all cells, then, the demagnetization fields are recalculated, and the process 
is repeated iteratively until a desired tolerance is reached.   A microscopic uniaxial anisotropy energy 
is also included, although using a strength that would be typical for Permalloy, it is almost irrelevant 
when compared to the exchange and demagnetization effects.   We have checked that this procedure gives
the same final states as integration of the Landau-Gilbert equations with damping.

The internal magnetic energy $E_{\rm int}$ of the particle is calculated.  This is the total magnetic energy
{\em minus} the interaction energy with the applied magnetic field, $-\vec\mu\cdot \vec{H}_{\rm ext}$.   
An applied magnetic field is used in the calculations to move the net magnetic moment around,
while it as well maps out the hysteresis loop.  In one set of simulations, the hysteresis loop was 
calculated with the applied field axis within the $xy$-plane at some angle $\phi_H$ to the $x$-axis.
There, the magnetization makes an angle $\phi_m$ to the $x$-axis.  Then the internal energy could be 
found as a function $E_{\rm int}(\phi_m)$, from which the anisotropy constant $K_1$ is determined,
by fitting to (\ref{potential}) in the form, 
\be
\label{pot1}
E_{\rm int}(\phi_m) = E_0+K_1 \sin^2\phi_m.
\ee
In another set of simulations, the applied field was set in the $xz$-plane, at some angle $\theta_H$ 
to the $x$-axis.   This tilts the net magnetic moment towards the $z$-axis by an angle $\theta_m$ from 
the $x$-axis.  Thus it gives $E_{\rm int}(\theta_m)$, which depends on both constants $K_1$ and $K_3$,
according to 
\be
\label{pot3}
E_{\rm int}(\theta_m) = E_0+(K_1+K_3) \sin^2\theta_m.
\ee
This allows anisotropy constant $K_3$ to be determined from the net stiffness, $K_{13}\equiv K_1+K_3$.   
It is important to note, that these potential 
functions $E_{\rm int}(\phi_m)$ and $E_{\rm int}(\theta_m)$ found this way do not depend on the 
particular angle chosen between the applied field and the $x$-axis.

In the following sections the Hamiltonian and algorithm is further specified.  Some details
about the demagnetization field calculation are given, especially concerning the Greens function.
Finally the results for elliptic particles are discussed.

%-------------------------------------------------------------------
\section{The particle model and its energetics}
%-------------------------------------------------------------------
We consider  thin elliptical particles with dimensions 
$L_x \times L_y \times L_z$,  where $L_x$ and $L_y$ refer to the major and minor diameters for
the elliptical particles.  The approach for a thin system has been presented in Ref. \cite{Wysin2010};
some of the main points towards finding the spatial structure of magnetization $\vec{M}(\vec{r})$ and
the particle's internal energy are summarized here.  

The system is partitioned into cells of size $a\times a\times L_z$
on a square lattice grid, where there is saturation magnetization $M_s$ within each cell.  Thus a selected 
cell $i$ has a magnetic moment ${\bf m}_i= M_s a^2 L_z \hat{m}_i$, that points in the direction of the
unit vector $\hat{m}_i$ and has magnitude $\mu_{\rm cell}=M_s a^2 L_z$.  There is only a single
layer of cells used, under the assumption that the perpendicular demagnetization effect leads to 
magnetization nearly independent of $z$, for the thin systems under consideration.  

The exchange interaction in continuum theory is taken in terms of the exchange stiffness $A$ 
(about 13 pJ/m for Py) as a volume integral,
\begin{equation}
\label{continuum-ex}
{\cal H}_{\rm ex} =  A \int dV ~ \nabla \hat{m} \cdot \nabla \hat{m}.
\end{equation}
where $\hat{m}=\vec{M}/M_s$ is the local reduced magnetization.  When expanded on the square lattice
of cells, this is equivalent to a nearest neighbor exchange term for the cells,
\begin{equation}
{\cal H}_{\rm ex} = -J \sum_{(i,j)} \hat{m}_i \cdot \hat{m}_j, \hskip 0.2in J = 2AL_z.
\end{equation}
A uniaxial anisotropy energy $K$ (about 100 J/m$^3$ for Py) is included as another volume integral
\begin{equation}
{\cal H}_{\rm uni} =  - K \int dV ~ (\hat{m} \cdot \hat{u} )^2 \rightarrow
- K a^2 L_z \sum_i \, (\hat{m}_i \cdot \hat{u} )^2 ,
\end{equation}
where the anisotropy axis here is taken as $\hat{u} = \hat{x}$.   The externally applied magnetic
field involves an energy of $-\vec{B}\cdot \vec\mu$ for any dipole, so 
\begin{equation}
{\cal H}_{\rm B} = -\int dV ~ \mu_0 \vec{H}_{\rm ext} \cdot \vec{M} \rightarrow
- \mu_0 M_s a^2 L_z \sum_i \, \vec{H}_{\rm ext}\cdot \hat{m}_i.
\end{equation}
Finally, the most important part of the interactions is the demagnetization field or dipolar
interaction.  Once the cells are defined on the grid with lattice spacing $a$, their dipole 
interaction could be described by a Hamiltonian,
\begin{equation}
\label{Hdd-micro}
{\cal H}_{\rm dd} = -\frac{\mu_0}{4\pi} \mu_{\rm cell}^2
\sum_{i>j} \frac{ \left[ 3(\hat{m}_i\cdot \hat{r}_{ij})(\hat{m}_j\cdot\hat{r}_{ij})
                                -\hat{m}_i\cdot \hat{m}_j \right]} 
{\left\vert \vec{r}_{i}-\vec{r}_{j}\right\vert^3} 
\end{equation}
However, this does not take into account the fact that the system is very thin.
The demagnetization field can be found very accurately for thin systems using a Greens 
function approach \cite{Huang03}.  To do that, instead we start from the continuum dipolar energy, 
\begin{equation}
{\cal H}_{\rm dd}= -\frac{1}{2} \mu_0 \int ~ dV ~ \vec{H}_M \cdot \vec{M}
\end{equation}
where $\vec{H}_M=-\vec\nabla\Phi_M$ is the demagnetization field at some point, and
$\Phi_M$ is its corresponding scalar magnetic potential. That field is produced by all the
dipoles, according to a Poisson equation for magneto-statics:
\begin{equation}
-\nabla^2\Phi_M = \rho_M, \quad {\rm where} \quad \rho_M = -\vec{\nabla}\cdot\vec{M}.
\end{equation}
Further, the discontinuity at the surfaces of the particle can be modeled as a magnetic surface
charge density, $\sigma_M= \vec{M}\cdot \hat{n}$, where $\hat{n}$ is the outward normal. In
particular,  that gives charge densities of opposite signs, $\sigma_M= \pm M_z$ on the upper and 
lower faces at $z=0, L_z$, respectively, under the assumption of uniform magnetization not
depending on $z$ within the cells.  The field of those surface charges is responsible for
keeping the magnetization close to the $xy$ plane.  There are also surface magnetic charges at
the edges of the island but those can be included into a localized volume charge for the
cells there.   But whether the magnetic charges are surface charges or volume charges 
makes no physical difference, however, and the solution of the Poisson equation is formally
\begin{equation}
\Phi(\vec{r}\,) = \int d^3r'~ \frac{\rho(\vec{r}\,')}{4\pi \left\vert \vec{r}-\vec{r}\,'\right\vert}  
\end{equation}
This can be used to find the demagnetization field at the point $\vec{r}=(x,y,z)$, and then
averaging that result over $z$ from $z=0$ to $z=L_z$.  The resulting demagnetization field
at a cell centered at $(x,y)$ has a vertical component $H_{M,z}$ and some in-plane component 
$\vec{H}_{M,xy}$. These are given by convolutions with appropriate 2D Green's functions,
involving only the in-plane position, denoted here as $\tilde{r}=(x,y)$.  For the vertical 
demagnetization component, one gets
\begin{equation}
\label{Hz}
{H}_{M,z}(\tilde{r}) 
= \int d^2 \tilde{r}\,' ~ G_z(\tilde{r}-\tilde{r}\,') ~ M_z(\tilde{r}\,')
, \quad \tilde{r} \equiv (x,y)
\end{equation}
\begin{equation}
\label{Gz}
G_z(\tilde{r}) = \frac{1}{2\pi L_z} \left[ \frac{1}{\sqrt{\tilde{r}^2+L_z^2}}
-\frac{1}{\left| \tilde{r} \right| } \right]
, \quad \tilde{r}^2 \equiv x^2+y^2
\end{equation}
For the in-plane components, there is
\begin{equation}
\label{Hxy}
\vec{H}_{M,xy}(\tilde{r})= \int d^2 \tilde{r}\,' ~ 
   \vec{G}_{xy}(\tilde{r}-\tilde{r}\,') ~ {\rho}(\tilde{r}\,') .
\end{equation}
\begin{equation}
\label{Gxy}
\vec{G}_{xy}(\tilde{r})= \frac{\hat{e}_{\tilde{r}}}{2\pi L_z} 
\left[ \sqrt{1+\left(\frac{L_z}{\vert\tilde{r}\vert}\right)^2}-1 \right]  .
\end{equation}
When applied, the unit vector $\hat{e}_{\tilde{r}-\tilde{r}\,'}$ points from source point $\tilde{r}\,'$
towards observation point $\tilde{r}$.   There is a singularity in $G_z(\tilde{r})$ as $\tilde{r}\rightarrow 0$, 
which is handled by averaging $G_z$ over a region with the same area as the cells being used, see 
Ref.\  \cite{Wysin2010} for further details on this averaging of the Green's functions.

Together with appropriate finite-difference approximations for the magnetic charge density, these expressions 
are used to get the demagnetization field.   The actual evaluation of these convolution integrals was performed
as multiplication in reciprocal space using a 2D fast Fourier transform approach \cite{Sasaki97}.  To simulate a 
free particle 
without the periodicity effects of the FFT (i.e., to avoid the wrap-around problem), the grid of the FFT is padded 
to a net size of $N_x\times N_y$, where $N_x$ and $N_y$ are the smallest powers of two satisfying $N_x \ge 2L_x/a$ 
and $N_y\ge 2L_x/a$.   Because we consider elongated particles, the calculations can be very fast due to $N_y$ 
being rather small compared to $N_x$, for high aspect ratio particles.  The FFT approach ends up giving the 
demagnetization field at the cell center positions (as well as at other points outside the particle, due to 
the padding).

It is convenient to measure magnetic fields $\vec{H}_{\rm ext}$ and $\vec{H}_{M}$ in units of 
the saturation magnetization $M_s$, just as done for the magnetization, $\hat{m}=\vec{M}/M_s$, 
and define the dimensionless fields,
\begin{equation}
\vec{h}_{\rm ext} \equiv \frac{\vec{H}_{\rm ext}}{M_s}, \hskip 0.3in 
\vec{h}_{M} \equiv \frac{\vec{H}_{M}}{M_s}.
\end{equation}
The basic (and largest) unit of energy is the exchange $J$ between neighboring cells.
Then the total effective Hamiltonian can be written in units of $J$ as
\begin{eqnarray}
\label{Hmm}
{\cal H} &+& -J \left\{ \sum_{(i,j)} \hat{m}_i \cdot \hat{m}_j \right. \\
&+& \left. \left( \frac{a}{\lambda_{\rm ex}} \right)^2 \sum_i \left[ \kappa (\hat{m}_i\cdot \hat{u})^2 +
\left( \frac{1}{2}\vec{h}_{M,i} + \vec{h}_{\rm ext} \right)\cdot \hat{m}_i \right] \right\} .
\nonumber
\end{eqnarray}
This is written in terms of the ferromagnetic exchange length $\lambda_{\rm ex}$ and the
scaled dimensionless uniaxial anisotropy $\kappa$, defined as
\begin{equation}
 \label{exl}
\lambda_{\rm ex} = \sqrt{\frac{2A}{\mu_0 M_S^2}}, \quad\quad
\kappa = \frac{K}{\mu_0\, M_s^2}.
\end{equation}
The magnetic internal energy $E_{\rm int}$ is of most interest.  That is this Hamiltonian, but
with the interaction with the external magnetic field (the last term) removed.

%--------------------------------------------------------------------------------------------------
For the calculations we used the values for Permalloy, $M_s\approx 860$ kA/m, $A\approx $ 13 pJ/m,
$K\approx 100$ J/m$^3$, then these give $\lambda_{\rm ex}\approx 5.3$ nm and 
$\kappa\approx 1.1\times 10^{-4}$.   Due to this small value of $\kappa$, in most of the 
calculations the intrinsic uniaxial anisotropy energy is negligible compared to the other 
energies of the system.  In most of the simulations the cell size was $a=2.0$ nm, except for
the smallest high aspect ratio particles, where values as low as $a=0.5$ nm  were used,
to produce a smoother edge to the particle.  These are sufficiently less than the exchange length 
to give a reliable description of the internal magnetic structure.

\begin{center}
\begin{figure}
\includegraphics[width=\smallfigwidth,angle=-90]{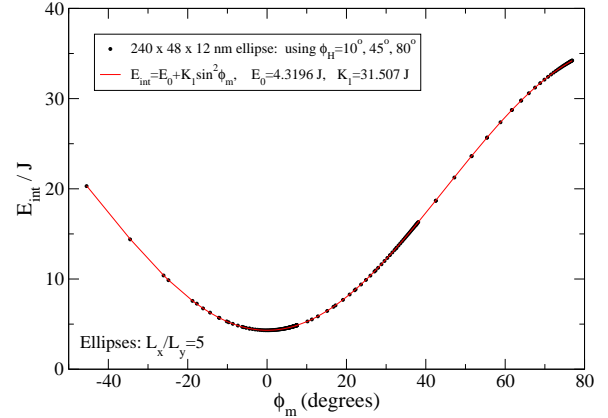}
\caption{\label{eint-phi} (Color online) The in-plane potential of an elliptical particle with a 5:1 aspect ratio, 
mapped out while determining the hysteresis loop (Fig. \protect\ref{mhB}).  The angle $\phi_m$ is 
the direction of the net particle moment $\vec\mu$ in the easy-plane.  The points come from 
simulations at the different indicated angles $\phi_H$ of the applied field from the long axis; 
all fall onto the same curve.  The fit gives a reliable estimate of anisotropy constant $K_1$.}
\end{figure}
\end{center}

%--------------------------------------------------------------------------------------------------
\section{Calculation procedures}
%--------------------------------------------------------------------------------------------------
The iteration procedure that moves the system towards the nearest local equilibrium is a local
spin relaxation algorithm \cite{Wysin96}, that points each cell's magnetic moment $\vec{m}_i$ (or its 
unit vector $\hat{m}_i$) along its local magnetic field $\vec{B}_i$.  That local field enters the 
undamped dynamic equation of motion,
\begin{equation}
\frac{d \vec{m}_i}{dt} = \gamma \vec{m}_i \times \vec{B}_i,
\end{equation}
and it is given by the variation of the Hamiltonian,
\begin{eqnarray}
\vec{B}_i &=& -\frac{\delta {\cal H}}{\delta \vec{m}_i} 
= -\frac{1}{\mu_{\rm cell}}\frac{\delta {\cal H}}{\delta \hat{m}_i}
= \frac{J}{\mu_{\rm cell}} \left\{ \sum_{\rm nbrs} \hat{m}_j \right.
\\
&+& \left. \left( \frac{a}{\lambda_{\rm ex}}\right)^2 
\left[2\kappa(\hat{m}_i\cdot\hat{u})\hat{u} + \frac{1}{2}\vec{h}_{M,i} + \vec{h}_{\rm ext} \right] \right\}  
\nonumber
\end{eqnarray}
Alternatively, this is the same as
\begin{eqnarray}
\vec{B}_i &=& \mu_0 M_s \left\{ \left(\frac{\lambda_{\rm ex}}{a}\right)^2 \sum_{\rm nbrs} \hat{m}_j  \right.
\\
&+& \left. \left[2\kappa(\hat{m}_i\cdot\hat{u})\hat{u} + \frac{1}{2}\vec{h}_{M,i} + \vec{h}_{\rm ext} \right] \right\}  
\nonumber
\end{eqnarray}
Either way, these define a unit vector along which to point the magnetic moment of cell $i$, 
$\hat{m}_i \rightarrow \hat{m}_i'$, where
\begin{equation}
\hat{m}_i' = \hat{b}_i = \frac{\vec{B}_i}{\vert \vec{B}_i \vert}.
\end{equation}
The alignment of $\hat{m}_i$ parallel to $\hat{b}_i$ is performed for every site of the grid, 
after which the demagnetization field must be recalculated.  The process moves the system towards 
lower energy. Each cell would stop moving if all became simultaneously aligned self-consistently with 
their local magnetic fields.  This does not insure a global energy minimum, however, and the
procedure does have the capability to generate the hysteresis loops.  The iteration is started from 
a partially aligned state of the cell dipoles, which are given some small random fluctuations away 
from perfect alignment.  For the hysteresis calculation, though, the last relaxed state at one applied
field is the initial state for the next value of applied field.

\begin{center}
\begin{figure}
\includegraphics[width=\smallfigwidth,angle=-90]{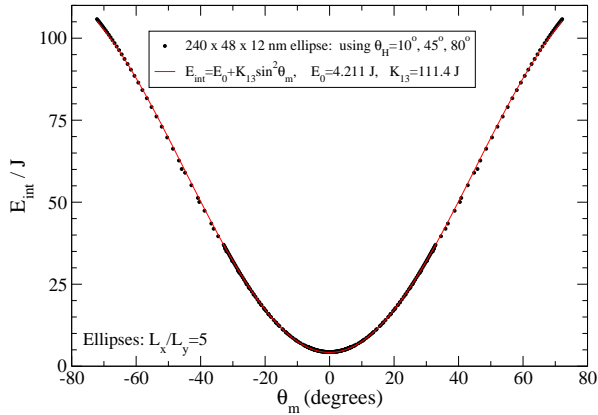}
\caption{\label{eint-theta} (Color online) The out-of-plane potential of the same elliptical particle with a 5:1 aspect 
ratio as in Fig. \protect\ref{eint-phi}, mapped out while determining the hysteresis loop (Fig.
\protect\ref{mhBz}).  
The angle $\theta_m$ is the tilting of the net particle moment $\vec\mu$ out of the easy-plane.  
The points from simulations at different angles $\theta_H$ of the applied field are combined into
one curve.  The fit gives a reliable estimate of the combined anisotropy constant $K_{13}=K_1+K_3$.}
\end{figure}
\end{center}

As the iteration proceeds, periodically (every 200 iterations of the system) the total magnetic moment 
$\vec\mu$ of the particle is calculated, by summing over the cell dipoles,
\begin{equation}
\vec\mu= (\mu_x, \mu_y, \mu_z) = \mu_{\rm cell} \sum_i \hat{m}_i 
\end{equation}
The iteration is stopped when the changes in any component of $\vec\mu$ are less than 1 part
in  $5 \times 10^7$ for two states separated by 200 iterations.  This is actually a more stringent 
stopping requirement than waiting for the energy to converge to the same precision.  

In one set of simulations, the applied magnetic field was directed within the $xy$-plane
at an angle $\phi_H$ to the $x$-axis [$\vec{H}_{\rm ext}=H_{\rm ext}(\cos\phi_H,\sin\phi_H,0)$].  
This results in the net magnetic moment $\vec\mu$ lying within the $xy$-plane, due to the strong 
perpendicular demagnetization.  The moment makes some angle $\phi_m<\phi_H$ to the $x$-axis, 
calculated from  $\phi_m= \tan^{-1}(\mu_y/\mu_x)$.  Thus, the in-plane potential energy function 
$E_{\rm int}(\phi_m)$ could be found, by scanning in applied magnetic field strength along the
chosen axis, and thereby calculating the hysteresis loop.  Then $K_1$ was found by fitting to the
form in (\ref{pot1}).

In the other set of simulations, the 
magnetic field was applied tilting out of the $xy$-plane, making an angle $\theta_H$ to the $x$-axis, 
that is, $\vec{H}_{\rm ext}=H_{\rm ext}(\cos\theta_H,0,\sin\theta_H)$.  This pulls $\vec\mu$ up an
angle $\theta_m<\theta_H$ from the easy ($xy$) plane, where $\theta_m= \tan^{-1}(\mu_z/\mu_x)$,
 and gives the opportunity to measure the potential $E_{\rm int}(\theta_m)$.   Again, the internal 
energy is calculated from the total Hamiltonian minus the interaction term with the applied external 
field. That energy was fitted to the out-of-plane potential (\ref{pot3}), whose stiffness is
due to the combination,  $K_{13} = K_1+K_3$.   Further, the potentials obtained did not depend on 
the choice of $\phi_H$ or $\theta_H$.    This could be seen by combining the internal energy curves for 
applied field at 45$^{\circ}$ and 80$^{\circ}$ to the $x$-axis. 

For the hysteresis curves, the total magnetic moment $\vec\mu$ was calculated, and normalized by the
particle volume $V$ to get the averaged magnetization inside the particle,  
$\langle \vec{M}\rangle = \frac{\vec\mu}{V}$.  Then the component of $\langle \vec{M}\rangle$
along the applied field axis is found,
\begin{equation}
\langle M_h \rangle \equiv \langle \vec{M}\rangle\cdot \hat{h}_{\rm ext}
\end{equation}
After scaling by the saturation magnetization, this is plotted versus the applied field magnitude 
also scaled by saturation magnetization ($h_{\rm ext}=H_{\rm ext}/M_s)$.

%--------------------------------------------------------------------------------------------------
\section{Results for elliptical particles}
%--------------------------------------------------------------------------------------------------

We considered thin elliptical particles with thicknesses all 1/20$^{\rm th}$ of the length, i.e., 
$g_3=L_x/L_z=20$, and aspect ratios $g_1=L_x/L_y = 3, 5$, and $8$. The lengths ranged from 120 nm to 
480 nm.  Some typical results for the internal energy curves are shown in Fig.\ \ref{eint-phi} for the
in-plane potential and Fig.\ \ref{eint-theta} for the out-of-plane potential of an elliptical
particle with $g_1=5$, with major axis 240 nm, minor axis 48 nm and thickness 12 nm.  The potentials
for in-plane motion of $\vec\mu$ fit very well to the functional form,
\begin{equation}
E_{\rm int}(\phi_m) = E_0 + K_1 \sin^2\phi_m.
\end{equation}
The constant $E_0$ is an irrelevant ground state energy when the particle is magnetized
along its long axis.
This same form also applies to the potential $E_{\rm int}(\theta_m)$, but with a coefficient $K_{13}=K_1+K_3$.
The fits are best for smaller particles, where the cells stay strongly aligned with each other,
and the reversal can be considered close to a uniform rotation process, for the most part. 
For the larger particles (length $> 400$ nm) this global alignment is lesser and the fits are good
but with considerably greater $\chi^2$.  Even so, the internal magnetization structure of the
relaxed states tends to be close to uniform.

%\vskip 0.2in
\begin{table}[ht]
\begin{center}
\begin{tabular}{|l|c|c|c|c|c|}
\hline
    & $L_x=120$ nm   & $L_x=240$ nm    & $L_x=480$ nm &   \\
\hline
$g_1=2$ & $K_1=6.35 J$  & $27.3 J$       & $111 J$      &   \\
\hline
$g_1=3$ & $7.32 J$      & $31.9 J$       & $134 J$      &   \\
\hline
$g_1=5$ & $6.96 J$      & $31.5 J$       & $133 J$      &   \\
\hline
$g_1=8$ & $7.39 J$      & $29.5 J$       & $118 J$      &   \\
\hline
\end{tabular}
\end{center}
\caption{\label{k1-table} Values of the in-plane anisotropy constant $K_1$ in units of $J=2AL_z$ 
for the indicated particle sizes and aspect ratios $g_1=L_x/L_y$. All of the particles calculated 
have $g_3=L_x/L_z=20$.}
\end{table}

%\vskip 0.2in
\begin{table}[ht]
\begin{center}
\begin{tabular}{|l|c|c|c|c|c|}
\hline
    & $L_x=120$ nm   & $L_x=240$ nm    & $L_x=480$ nm &   \\
\hline
$g_1=2$ & $K_{13}=79.0 J$ & $314 J$        & $1250 J$     &   \\
\hline
$g_1=3$ & $50.7 J$     & $201 J$        & $804 J$      &   \\
\hline
$g_1=5$ & $28.1 J$     & $111 J$        & $444 J$      &   \\
\hline
$g_1=8$ & $15.7 J$     & $62.6 J$       & $250 J$      &   \\
\hline
\end{tabular}
\end{center}
\caption{\label{k13-table} Values of the combined anisotropy constant $K_{13}$ in units of $J=2AL_z$ 
for the indicated particle sizes and aspect ratios $g_1=L_x/L_y$. All of the particles calculated have 
$g_3=L_x/L_z=20$.}
\end{table}

\begin{center}
\begin{figure}
\includegraphics[width=\smallfigwidth,angle=-90]{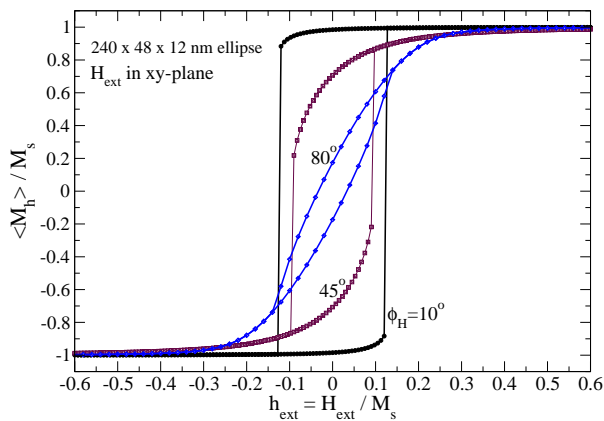}
\caption{\label{mhB} (Color online) Hysteresis loops for an elliptical particle as in Fig. \protect\ref{eint-phi}
with an in-plane applied field at the indicated angles $\phi_H$ to the long axis of the particle.}
\end{figure}
\end{center}

\begin{figure}
\includegraphics[width=\smallfigwidth,angle=-90]{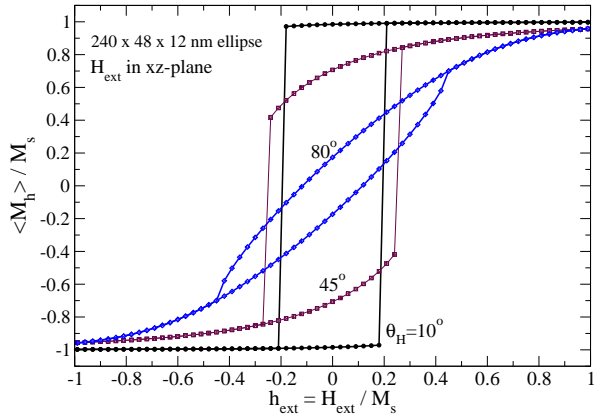}
\caption{\label{mhBz} (Color online) Hysteresis loops for an elliptical particle as in Fig. \protect\ref{eint-theta} 
with the applied field tilted out of the $xy$-plane at the indicated angles $\theta_H$ from the 
long axis of the particle.}
\end{figure}

Fitting results are summarized in Tables \ref{k1-table} and \ref{k13-table}, with the constants presented
in units of $J=2AL_z$.  The hard-axis anisotropy constant $K_3$ derived from those results is 
presented in Table \ref{k3-table}.   The  constant $K_3$ is consistently stronger 
than the easy-axis constant $K_1$, as to be expected from the greater surface area of the lower and upper faces 
at $z=0, L_z$, compared to the very limited surface area of the edge of the ellipse.  The energy unit 
$J$ itself varies according to the thickness.  Thus it makes sense to also look at results for the
constants in joules.  

%\vskip 0.2in
\begin{table}[ht]
\begin{center}
\begin{tabular}{|l|c|c|c|c|c|}
\hline
    & $L_x=120$ nm   & $L_x=240$ nm    & $L_x=480$ nm &   \\
\hline
$g_1=2$ & $K_3=72.7 J$ & $287 J$        & $1140 J$     &   \\
\hline
$g_1=3$ & $43.4 J$     & $169 J$        & $670 J$      &   \\
\hline
$g_1=5$ & $21.1 J$     & $79.9 J$        & $311 J$      &   \\
\hline
$g_1=8$ & $8.30 J$     & $33.1 J$       & $132 J$      &   \\
\hline
\end{tabular}
\end{center}
\caption{\label{k3-table} Values of the hard-axis anisotropy constant $K_3$ in units of $J=2AL_z$
for the indicated particle sizes and aspect ratios $g_1=L_x/L_y$. All of the particles calculated have
$g_3=L_x/L_z=20$.}
\end{table}

Generally, $K_3/J$ increases proportional to the area of the ellipse, $\frac{1}{4}\pi L_x L_y$, 
multiplied by the thickness $l_z$, so that in fact $K_3$ (in joules) is linearly proportional to the volume 
of the particles.  Also, one sees that $K_3$ decreases with increasing aspect ratio for particles of 
the same length; this is because the particle volume is decreasing. 
On the other hand, $K_1/J$ depends very weakly on the aspect ratio for the particle sizes tested.  
In addition, the calculations suggest that $K_1$ increases somewhat faster than the particle volume.  
The weak dependence of $K_1$ on the shape of the ellipse (at these larger values of $g_1$) is surprising.

\begin{figure}
\includegraphics[width=\smallfigwidth,angle=-90]{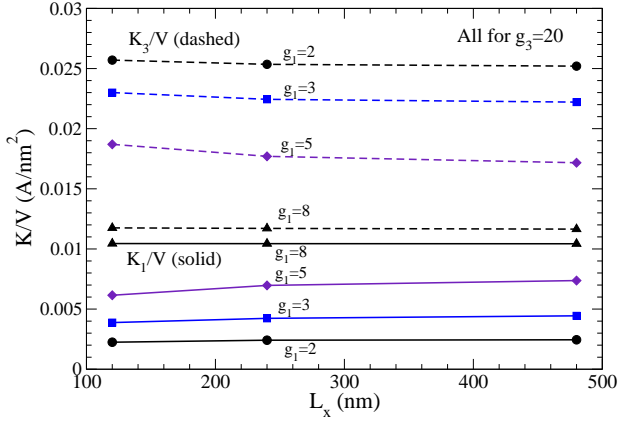}
\caption{\label{k1k3-fig} (Color online) The anisotropy constants $K_1$ (solid curves) and $K_3$ (dashed curves) 
scaled by elliptical particle volume, versus particle lengths, for the indicated $g_1$ aspect ratios. 
All data has $g_3=20$.  The values of $K/V$ are given in units of $A$/nm$^2$, where $A$ is the 
exchange stiffness.  $K_1/V$ increases with aspect ratio while $K_3/V$ decreases, and they become equal
at high aspect ratio.}
\end{figure}

To clarify the results we also show the constants converted to energy densities, 
both $K_1/V$ and $K_3/V$ in joules/nm$^3$, in Fig. \ref{k1k3-fig}.  The actual units are the
exchange stiffness $A$ (units of joules/nm) divided by squared nanometers.  One finds very little
dependence of either energy density constant, $K/V$, on the particle size, however, again it
is clear that $K_3$ is always larger  than $K_1$.  Furthermore,  the easy-axis anisotropy constant $K_1/V$
does increase rapidly with the in-plane aspect ratio $g_1$, and the relation could be close
to a linear relationship.  Although the values of $K_3/V$ are always greater than the corresponding
$K_1/V$, these hard-axis energy densities $K_3/V$ decrease slightly with increasing aspect ratio $g_1$.
At large aspect ratio, the two constants become nearly the same, which would have to be the
case for a needle-shaped magnet.

%-------------------------------------------------------------------------------------------------
\subsection{The magnetization structure}
%-------------------------------------------------------------------------------------------------
In the high aspect ratio particles, the magnetization states are very close to uniform, even
when undergoing the reversal. The elongated particle has such a strong anisotropic effect that
the magnetization cells move almost in a synchronized motion.  For particles with smaller
aspect ratio, one starts to see some weak variations in the magnetization inside the particle.

\begin{figure}
\includegraphics[width=\tinyfigwidth,angle=0]{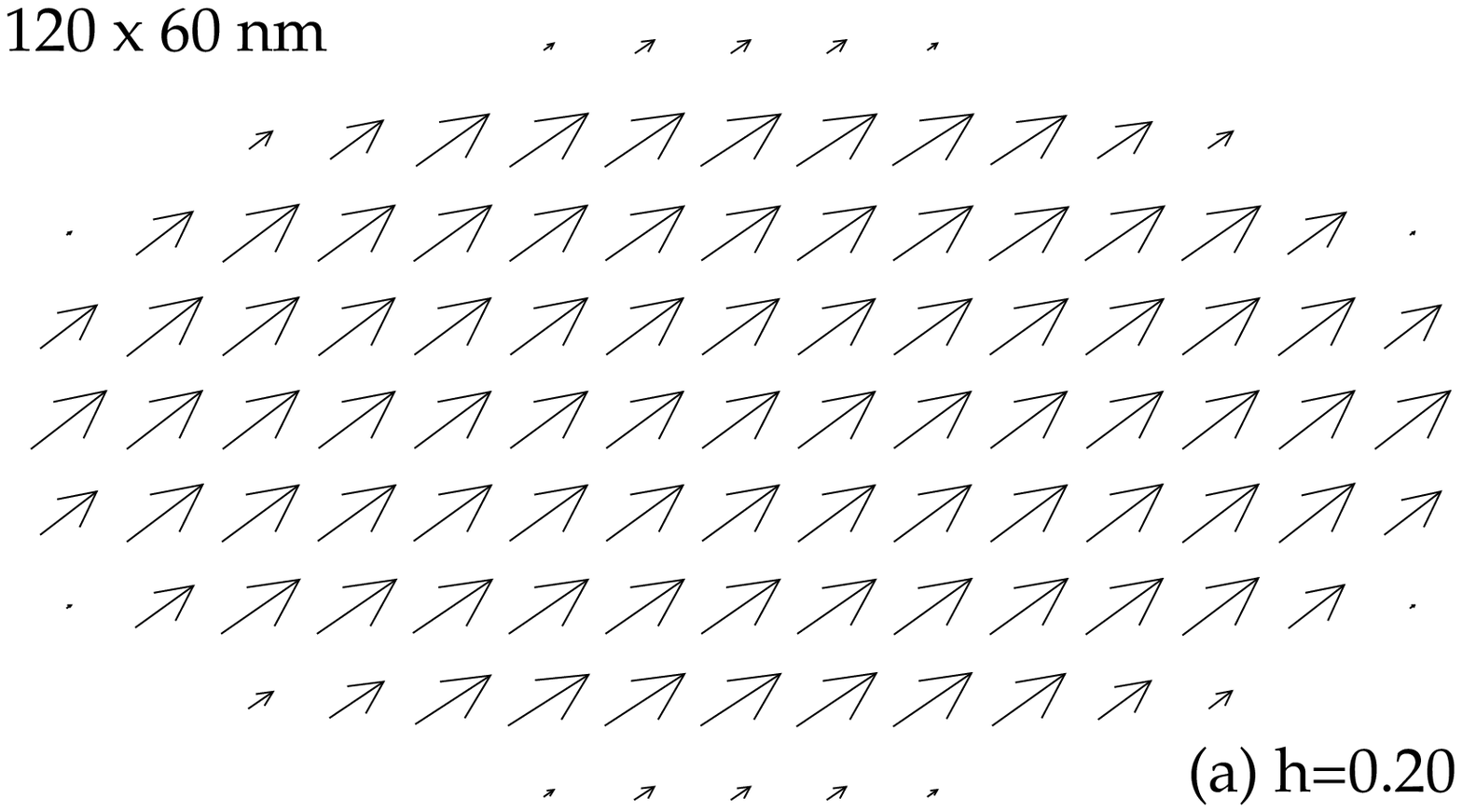}
\includegraphics[width=\tinyfigwidth,angle=0]{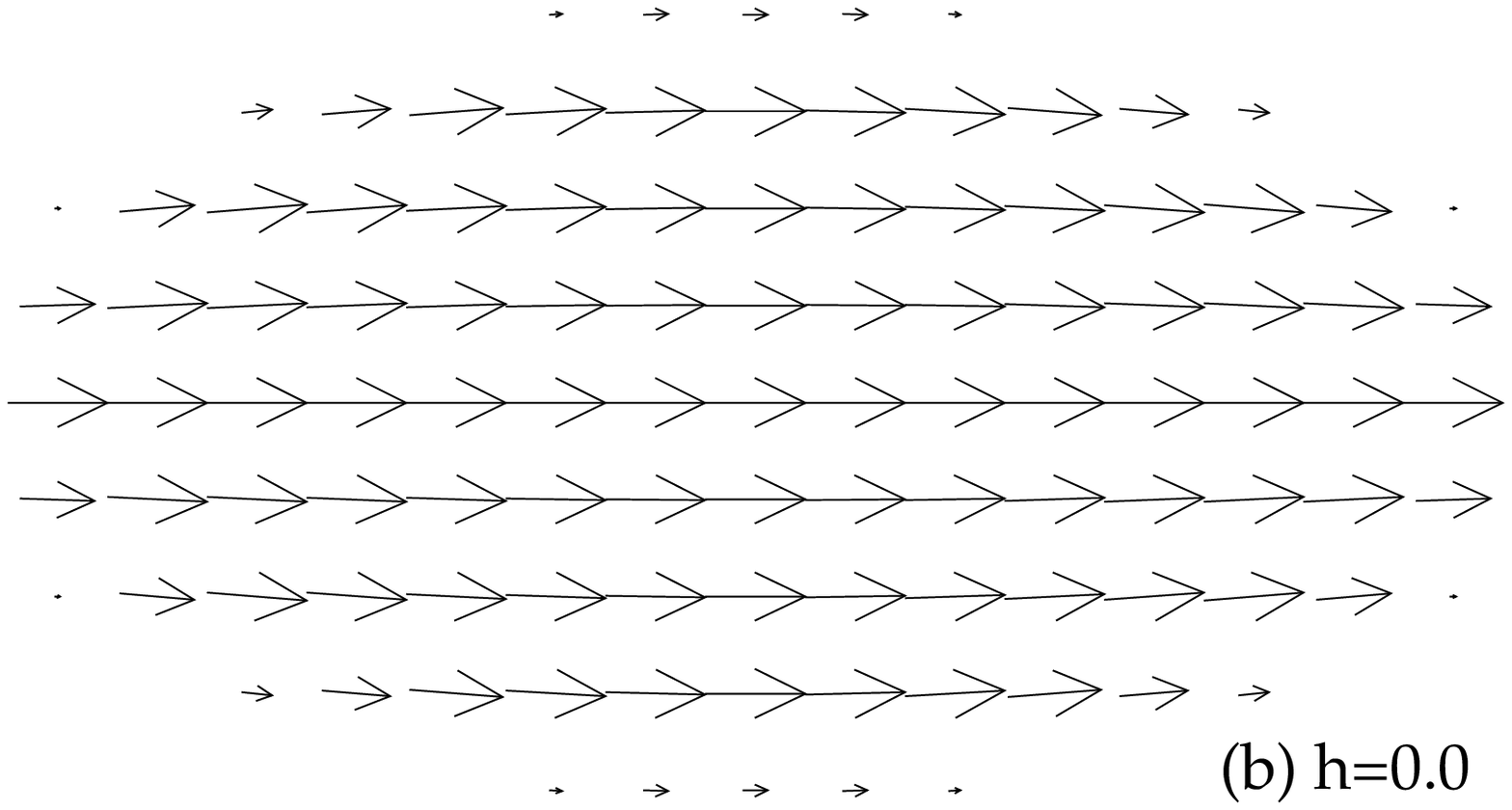}
\includegraphics[width=\tinyfigwidth,angle=0]{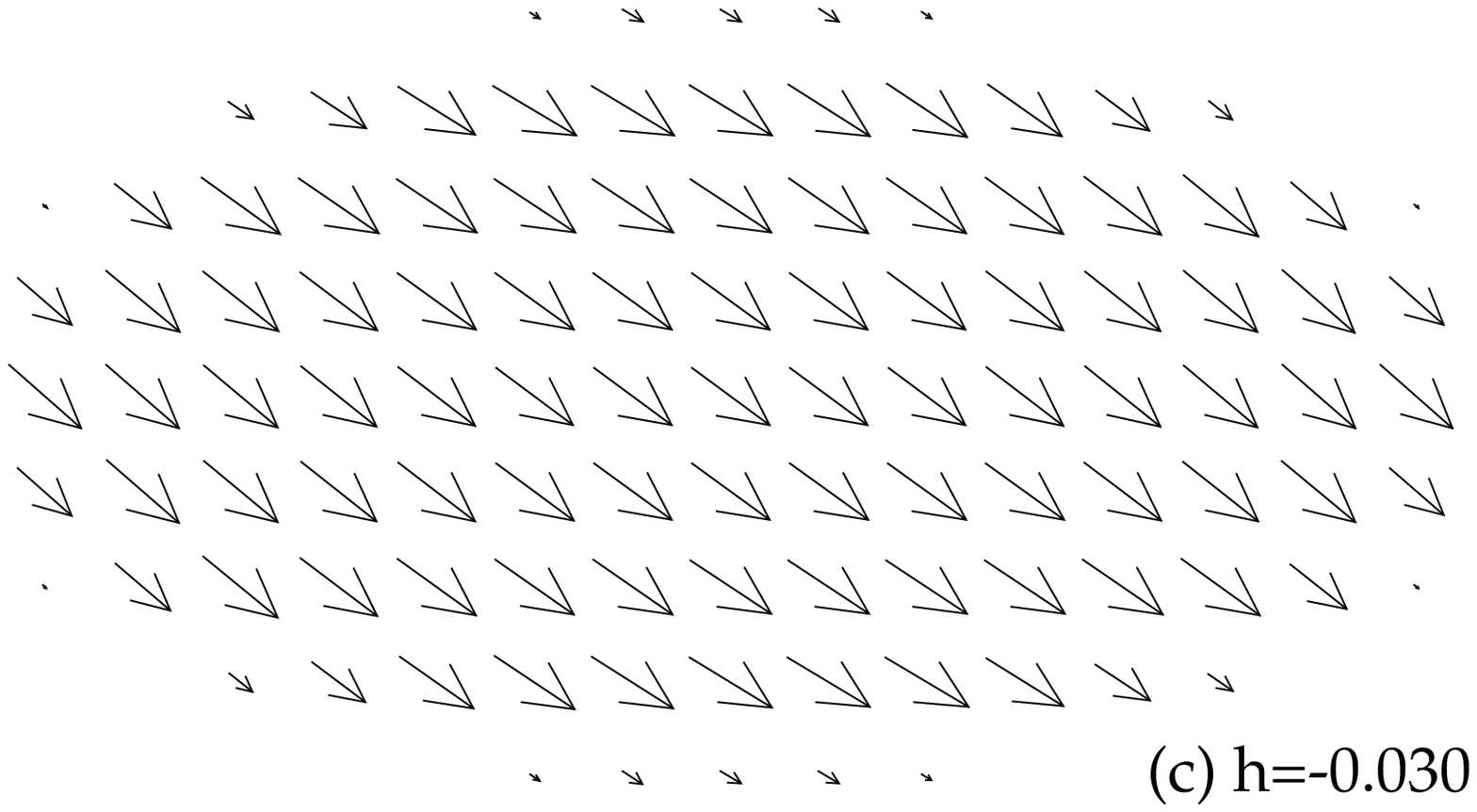}
\includegraphics[width=\tinyfigwidth,angle=0]{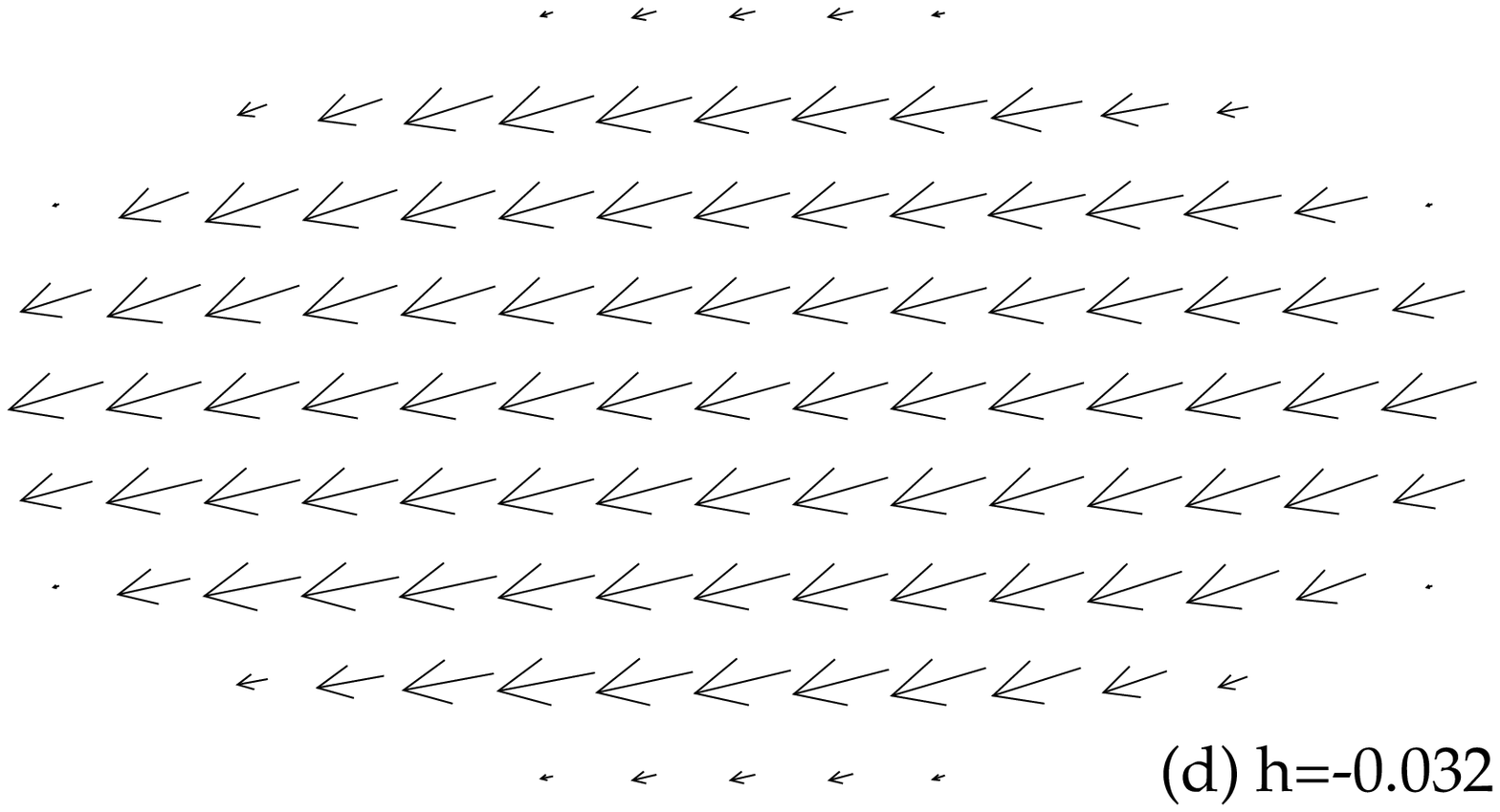}
\caption{\label{sp2A}  Magnetic configurations for a $120 \times 60 \times 6$ nm particle with
magnetic field applied at +45$^{\circ}$ above a horizontal axis pointing to the right.  
The arrows are the coarse-grained averages of $3 \times 3$ groups of cells.  In (a), the
external field is $h=0.20$; in (b) $h=0.0$;  (c) $h=-0.030$, just before reversal; (d) $h=-0.032$,
just after reversal.}
\end{figure}

To get an idea of the size of this effect, some configurations are presented for ellipses
with $g_1=2$,  which has the strongest effect of all the particle shapes presented earlier.
In Fig.\ \ref{sp2A} some configurations are shown for a $120 \times 60 \times 6$ nm particle,
at different applied field strengths 45$^{\circ}$ to the particle's long ($+x$) axis. The
points shown are at (a) close to saturation, (b) zero applied field, (c) a negative field
close to reversal, and (d) a negative field just after reversal.  For the most part, the
magnetization stays nearly uniform for this relatively small particle.

\begin{figure}
\includegraphics[width=\tinyfigwidth,angle=0]{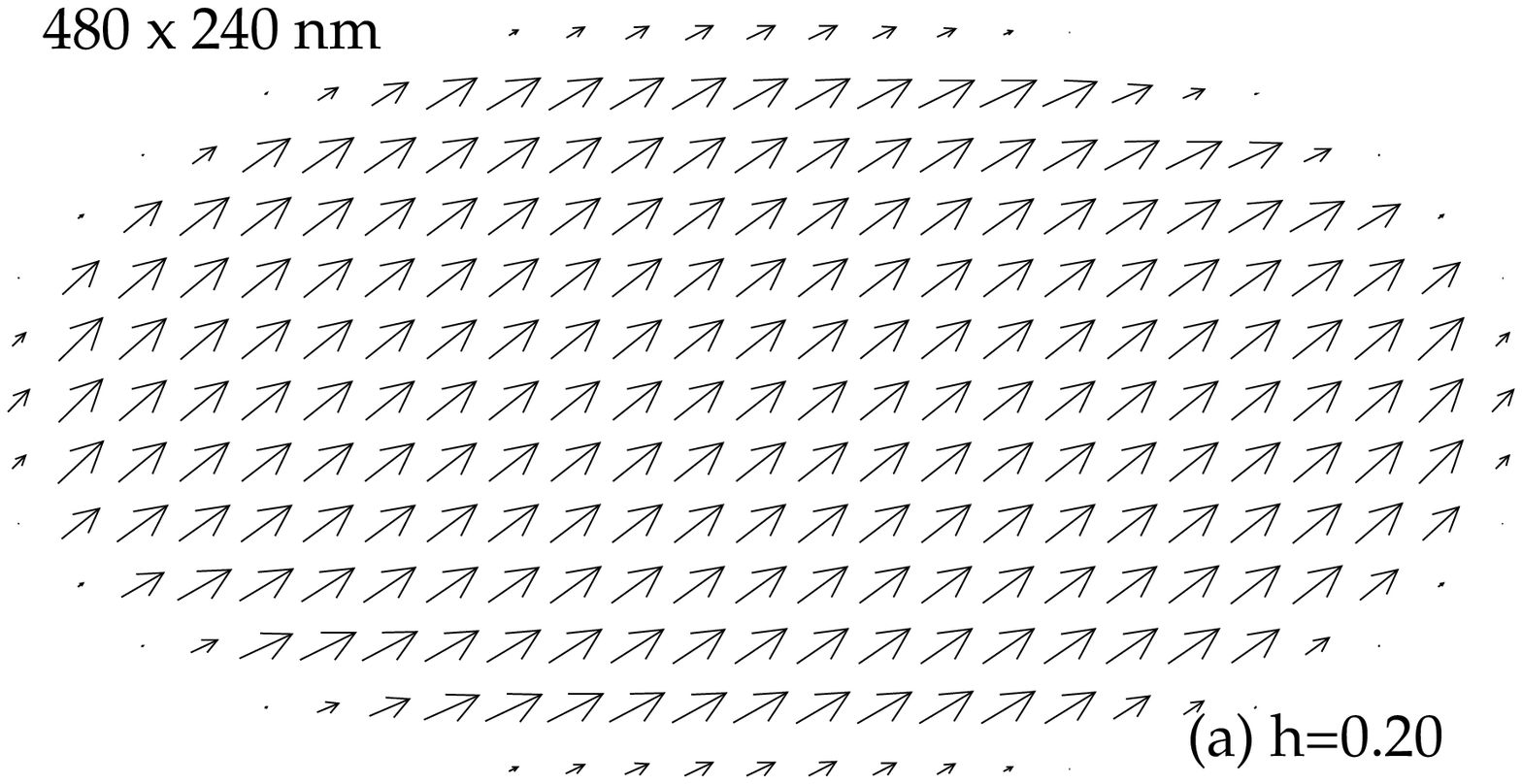}
\includegraphics[width=\tinyfigwidth,angle=0]{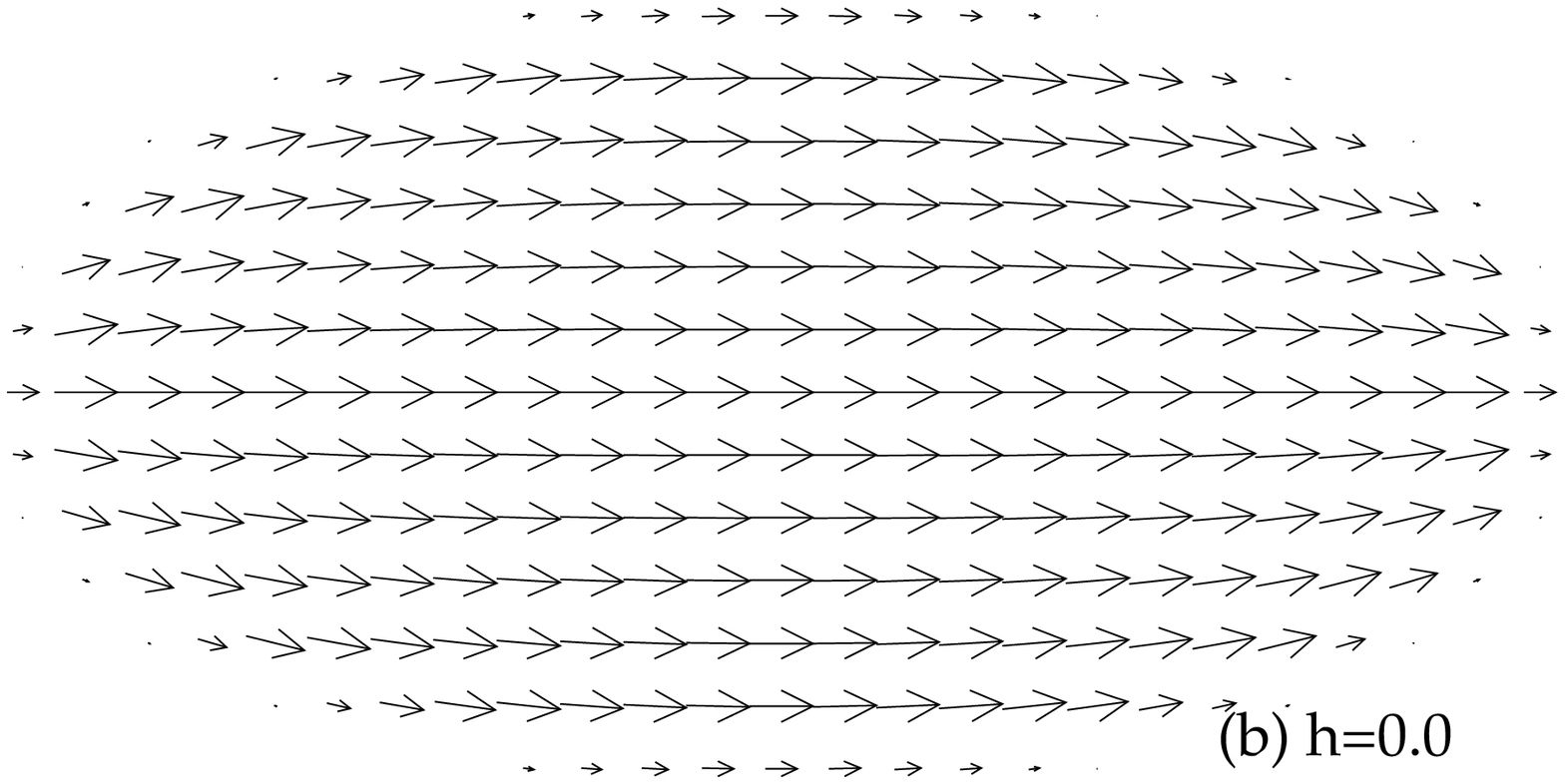}
\includegraphics[width=\tinyfigwidth,angle=0]{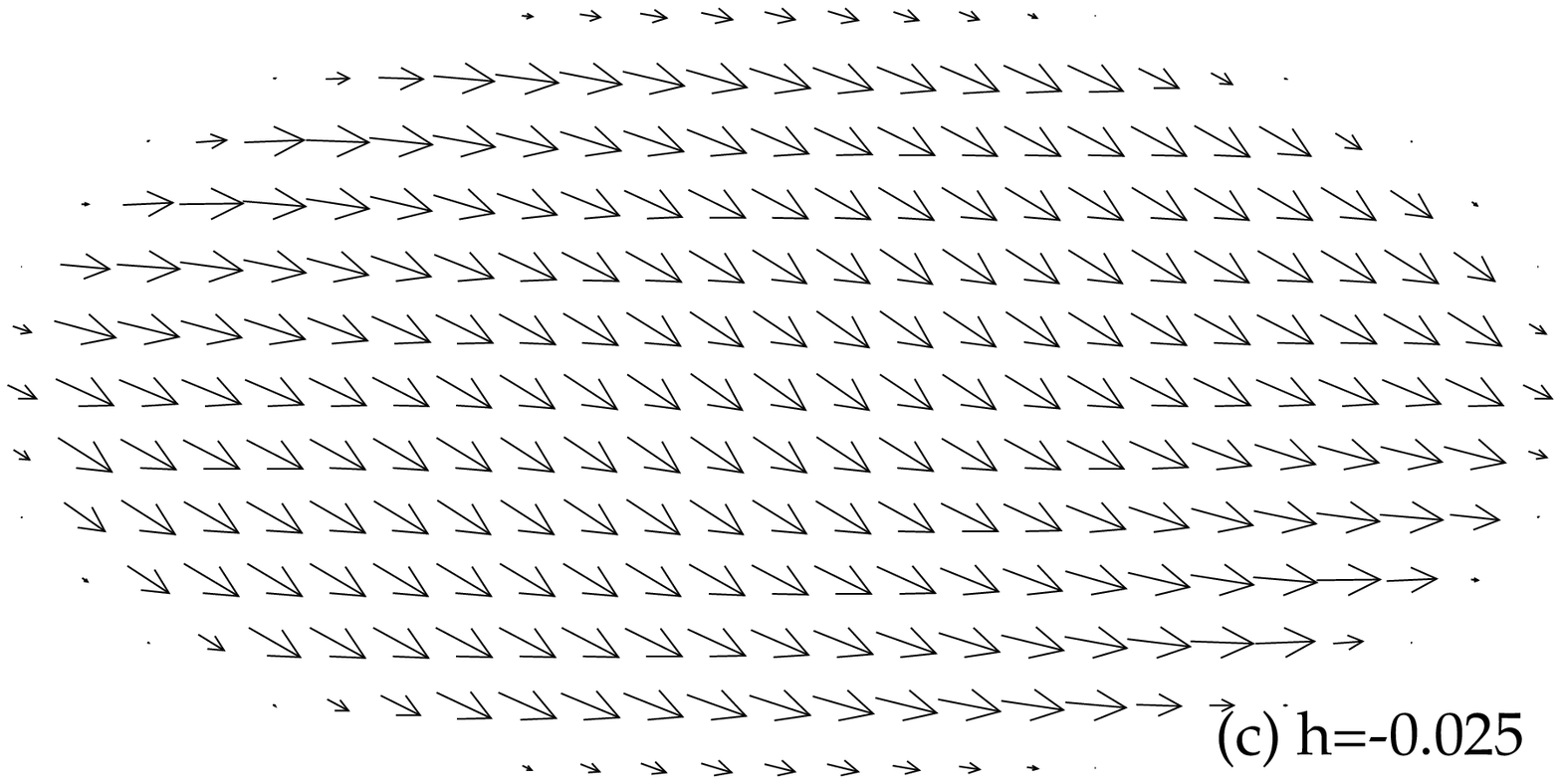}
\includegraphics[width=\tinyfigwidth,angle=0]{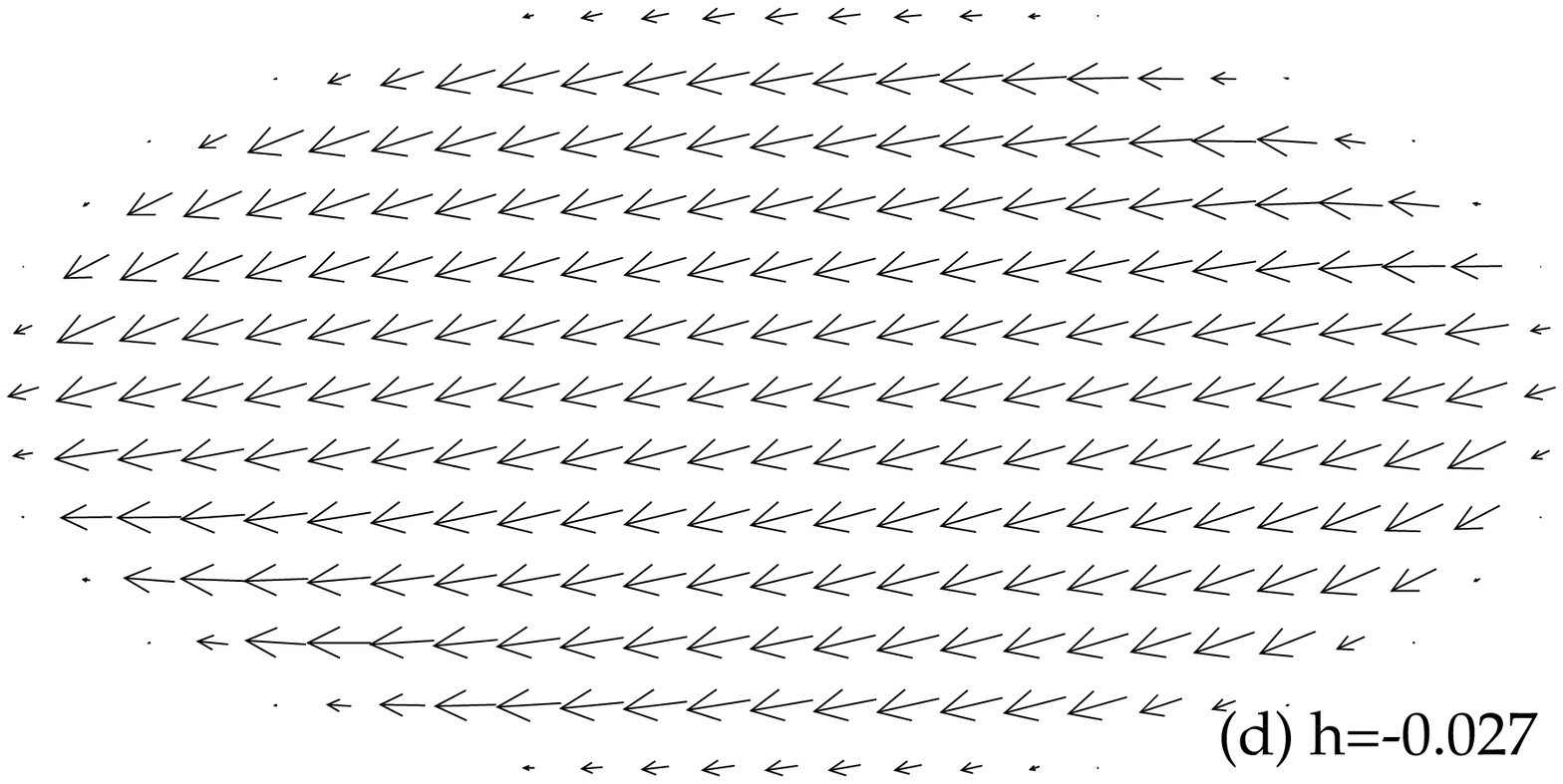}
\caption{\label{sp2C}  Magnetic configurations for a $480 \times 240 \times 24$ nm particle with
magnetic field applied at +45$^{\circ}$ above a horizontal axis pointing to the right.  
The arrows are the coarse-grained averages of $9 \times 9$ groups of cells.  In (a), the
external field is $h=0.20$; in (b) $h=0.0$;  (c) $h=-0.025$, just before reversal; (d) $h=-0.027$,
just after reversal. Note the enhanced curvature of the field compared to that in the smaller
particle in Fig.\ \protect\ref{sp2A}.}
\end{figure}

Another example is presented in Fig.\ \ref{sp2C}, like the first example, but $2 \times$ 
larger in all three dimensions.  The four configurations shown correspond to the same four
types of states as presented for the smaller particle.  The main difference here is that
a nonuniform magnetic structure develops.  At zero field, the structure points inward/outward
towards the poles on the long axis.  For the configurations just before and after reversal,
a wave-like structure is present.  These spatial variations are due to the dipolar interactions;
in even lower aspect ratio particles ($g_1<2$), they lead to C-states and even vortices entering 
the particle.

%---------------------------------------------------------------------
\subsection{Particles with lower aspect ratio $g_1<2$}
%---------------------------------------------------------------------
When $g_1\rightarrow 1$, the ellipse becomes circular and the easy-axis anisotropy must vanish.
Using smaller $g_1$ is a way to produce particles with weaker easy-axis anisotropy constant.
However, as the system becomes closer to circular, the lowest energy configuration, especially
near zero applied magnetic field,  tends to be nonuniform.  The ground state can tend towards 
a C-state or a vortex state if the particle is of sufficient size. The above results do not 
apply to that situation, especially because the nonuniform magnetization cannot be mapped into 
the model of an individual magnetic moment moving in an effective potential. 

To verify this, some particles were also calculated at small ellipticity, where $K_1\approx 0$,
using $g_1=1.25$ and $g_1=1.11$.  Generally, at these ratios, if there was a stable single-domain ground 
state (for smaller particles only),  the tendency is for the moments to try to follow the border,
and point inwards or outwards from the poles at the long ends.  At larger particle size this tilting
eventually moves the system irreversibly to a vortex ground state.  Until the vortex state is 
reached,  an effective potential can be estimated, however, from the practical point of view it 
may be of limited use.

%---------------------------------------------------------------------------
\subsection{Thicker particles}
%---------------------------------------------------------------------------
\begin{figure}
\includegraphics[width=\smallfigwidth,angle=-90]{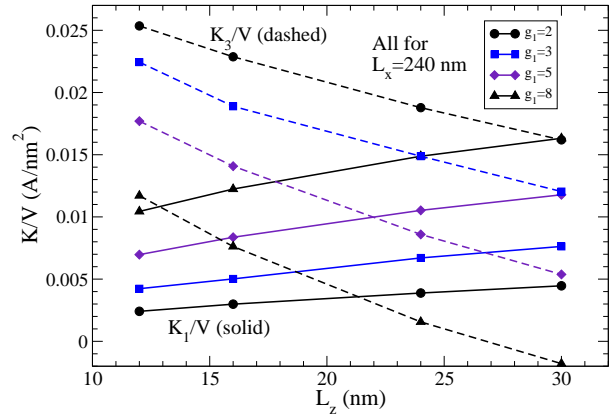}
\caption{\label{k1k3Lz-fig} (Color online) The anisotropy constants $K_1$ (solid curves) and $K_3$ (dashed curves)
scaled by elliptical particle volume, versus particle thicknesses, for the indicated $g_1$ aspect ratios.
All the data is for particles of length $L_x=240$ nm.  The $K_3/V$ constant crosses below zero for the
thickest high-aspect-ratio particles, which have become needle-like and no longer satisfy the
assumption of a thin particle. That is the case of a particle with only uniaxial anisotropy.}
\end{figure}

The particles with $g_3=20$ can be too thin to hold a magnetic moment stable against
room-temperature thermal fluctuations.  Thus it is important to consider the changes
when thicker particles are used.    Further calculations were carried out for 240 nm long
particles to get  results for $g_3=20, 15, 10,$ and $8$, corresponding to thicknesses of 
12, 16, 24 and 30 nm, respectively.  
The results for $K_1/V$ and $K_3/V$ are shown in Fig.\ \ref{k1k3Lz-fig}.  As could be expected, 
the thicker particles have weaker out-of-plane anisotropy $K_3/V$, while $K_1/V$ 
increases due to the thicker lateral edges, but at a rate less than linear
in the thickness.    We expect that these per-volume energy constants have only very
weak dependence on the particle length, as was already seen in the results presented
above for 12.0 nm thickness.  

%---------------------------------------------------------------------------
\section{Conclusions and Discussion}
%---------------------------------------------------------------------------
The anisotropy properties of thin elliptical ferromagnetic particles have been estimated, 
based on a 2D micromagnetics model that employs Green's functions for the calculation of 
the demagnetization fields.  For the high-aspect-ratio particles being considered,  the 
magnetization was found to be close to uniform inside the particles.  Then it was possible 
to map out the changes in the internal energy versus the direction of the net magnetic moment 
$\vec\mu$, which itself acts as a collective coordinate.  The typical particles tend to have 
stronger anisotropy in the hard-axis direction ($K_3/V$) than in the easy-axis direction 
($K_1/V$), however, these two energy scales approach each other for needle-like 
particles, as expected.   The results could be of practical application in the design
and analysis of artificial spin-ice with desired dynamics, beyond the usual Ising energetics.  

In the theoretical study of artificial spin ice materials, it is usual to
replace the islands by point-like dipoles with an Ising-like behavior.
Indeed, all theoretical calculations for the properties of these systems
were obtained with this approach. However, a more realistic description of
these artificial spin ices should require models beyond the Ising
approximation, such as continuous magnetic moments with anisotropy considered
in this work. In such a case, although the main properties of a spin ice system may
not undergo strong alterations, several quantities would change their
values. For instance, a recent work about the thermodynamics of the square
lattice \cite{Silva+11} has suggested a possible phase transition in this system,
occurring at a temperature of $7.2D$, where $D$ is the coupling constant of the
dipolar interaction among the islands. Of course, the transition temperature or 
similar quantities should be dependent on the island sizes and anisotropies, 
but this dependence cannot
be perceived with the Ising approach. It is very probable that the correct
critical temperature must be much smaller than $7.2D$ since the total magnetic
moment of an island has more degrees of freedom, and effectively moves in
a softer potential. In addition, the properties
must also be dependent on the islands' shapes, etc. So, the results obtained
here are of fundamental importance for developing this field not only
theoretically but also experimentally, suggesting protocols for improving
experiments,  and including studies about their dynamics.

%---------------------------------------------------------------------------
\section{Acknowledgments}
%---------------------------------------------------------------------------
G.M. Wysin appreciates the hospitality of The Department of Physics at Universidade
Federal de Vi\c cosa where this work was carried out, and is grateful for the financial 
support of FAPEMIG grant BPV-00046-11 for visiting research professor while at UFV.

\end{document}